# On two approaches to studying the aftershocks of a strong earthquake

A,V. Guglielmi

*Schmidt Institute of Physics of the Earth, Russian Academy of Sciences, 10-1 Bolshaya Gruzinskaya street,123242 Moscow, Russian Federation*

*E-mail*: *guglielmi@mail.ru*

The paper is devoted to comparing two approaches to the study of aftershocks. The methodological foundations of the traditional approach were laid many years ago. A new approach has emerged relatively recently. The two approaches differ from each other in the object, purpose and method of research. The differences are as follows. With the new approach, attention is focused not on aftershocks, but on the source of the earthquake. The evolution of the source is studied experimentally, and not the degradation of the frequency of aftershocks. Instead of a speculative selection of empirical formulas, the source deactivation coefficient is measured, variations in the coefficient are observed, and only on the basis of measurements and observations are conclusions drawn about the dynamics of the source. Thus, the divergence between the two approaches is doctrinal. The new approach turned out to be effective. Through targeted analysis of aftershock data, the Omori epoch and the phenomenon of bifurcation of the earthquake source were discovered. The purpose of further research is indicated.

## 1. Introduction

130 years ago, Omori found that the frequency of aftershocks decreases hyperbolically with time [1]. Omori's law has the following form

$$n(t) = \frac{k}{c+t}. \qquad (1)$$

Here $n$ is the frequency of aftershocks, $t \geq 0$, $k > 0$, $c > 0$. The parameter $k$ characterizes the earthquake source as a physical system. Omori's law is one-parameter, since the value of $c$ is determined by the choice of the time reference point.

100 years ago, Hirano [2] proposed changing Omori law with a power law of the form

$$n(t) = \frac{k}{(c+t)^p}. \qquad (2)$$

Here $p > 0$ is an additional fitting parameter. In other words, the power law is two-parameter. When $p = 1$, formula (2) coincides with Omori formula (1).

In the second half of the last century, Utsu [3–5] showed the effectiveness of formula (2) in the analytical description of aftershocks. Since then, formula (2) has been widely used in seismology for processing and analyzing aftershocks (see, for example, [6–11]). Sometimes formula (2) is called the Utsu law [11], although it would be more correct to call it the Hirano–Utsu law. In the course of research, it was found that the exponent is on average $p = 1.1.$, but varies from case to case within a wide range (approximately from 0.7 to 1.5).

The described approach to the problem is based on a certain methodological setting. Research methods generally change over time, and usually for the better. In recent years, a new approach to the problem has begun to emerge, epistemologically different from that of Omori, Hirano and Utsu. The difference concerns the object, purpose and method of research [12–16]. This paper is devoted to the description and comparison of two approaches to the study of aftershocks.

## 2. Selection of empirical formulas

The search for an empirical formula approximating observational data is widely used in the practice of studying natural phenomena. With a good choice, the formula may turn out to be fundamental. We do not know for what reasons Omori chose the hyperbolic function to formulate the law of aftershock evolution. (To get ahead of ourselves, let's say that the choice turned out to be relatively successful.) It is quite possible that considerations of simplicity played no small role for him. At the same time, as Dirac noted, the beauty of the mathematical formulation of the theory may be even more important than compliance with experimental data. And the hyperbola, being one of the conic sections, is extremely attractive in its own way.

It is more difficult to understand why Hirano and Utsu preferred the power law. Of course, the two-parameter formula (2) better approximates the experimental points than the one-parameter formula (1). But this improvement comes at a high price. Indeed, if in the Omori law the parameter $k$ is dimensionless, then in the Hirano-Utsu law the parameter $k$ has no specific dimension at all. This deprives formula (2) of physical meaning. The dimension $k$ depends on the value $p$. For example, when $p = 1.1$ we have $[k] = s^{1/10}$. This is unacceptable from the point of view of physical-mathematical spelling. It is impossible to rationally explain the century-old misconception on this score. This is an amazing thing. Even the famous mathematician and geophysicist Harold Jeffreys used (2) to approximate the flow of aftershocks after the 1927 earthquake in Tango (Japan) [17]. Meanwhile, formula (2) does not have the status of a law, since in physics the phenomenological parameters have a fixed dimension.

There is another, less significant drawback of the Hirano-Utsu law. From (2) it follows that $\lim n(t) = 0$ at $t \to \infty$, while observational experience tells us that the specified limit is non-zero. In asymptotics, the source goes into the background seismicity regime. The frequency of tremors $n_\infty$, generally speaking, is different from zero. In the next section of the paper we will present a two-parameter law of evolution, devoid of the disadvantages of the Hirano–Utsu law mentioned here.

### 3. Logistic law of aftershock evolution

If, in the mathematical formulation of the law, we use two phenomenological parameters, then a vast scope for choice opens up. In order not to wander in the dark, we need to be guided by some rational considerations. We will look for a curve for the decline in the frequency of aftershocks among the curves that have long been well known in science. Let us also impose the conditions of simplicity and beauty in the Dirac sense. This particular choice is not completely determined, but logistic curves immediately come to mind. They have long been successfully used in biology, chemistry, demography and other sciences. Let us show that one of the two classes of logistic curves is perfectly suitable for solving our problem [18].

Logistic curves are solutions to the logistic equation

$$\frac{dn}{dt} = n(\gamma - \sigma n). \qquad (3)$$

Here $\gamma$ and $\sigma$ are two parameters characterizing the earthquake source. The problem of dimensionality of parameters does not arise for us, and we immediately solve the problem of the asymptotic behavior of aftershocks: $n \to n_\infty = \gamma / \sigma$ at $t \to \infty$.

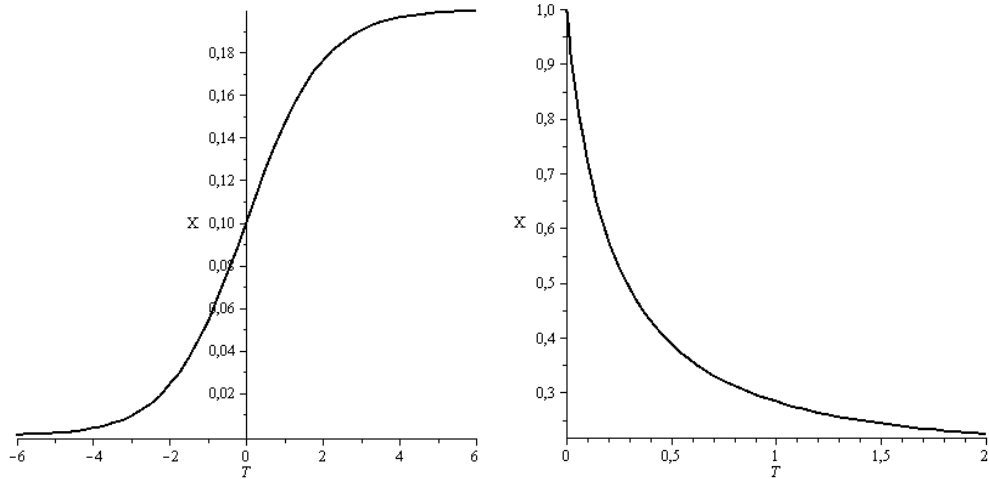

**Fig. 1.** Two classes of solutions to the logistic equation of evolution. The dimensionless quantity $T = \gamma t$ ( $X = n/n_\infty$ ) is plotted along the horizontal (vertical) axis.

There are two classes of solutions to the logistic equation – rising and falling solutions (Figure 1). The choice between classes occurs when setting up the Cauchy problem for equation (3). Let us explain this procedure.

We will set the initial condition $n(0) = n_0$ at $t = 0$, and we will look for a solution at $t > 0$. Increasing logistic curves arise when $n_0 < n_\infty$. They have long been widely used in biology, chemistry and sociology, as mentioned above. Decreasing logistic curves arise at $n_0 > n_\infty$. They are the ones that are of interest to us.

The curve in the right panel of Figure 1 is remarkably similar to an Omori hyperbola. Moreover, when $n_0 \gg n_\infty$, the upper segment of the logistic curve is practically indistinguishable from a hyperbola. Let's move on to investigating this issue.

When $n_0 > n(t) \gg n_\infty$ we can neglect the first term on the right side of equation (3):

$$\frac{dn}{dt} + \sigma n^2 = 0. \qquad (4)$$

Let us show that the shortened differential evolution equation (4) is equivalent to the hyperbolic Omori law [18].

The solution to the Cauchy problem for equation (4) has the form

$$n(\tau) = \frac{n_0}{1 + n_0 \tau}, \qquad (5)$$

where

$$\tau = \int_0^t \sigma(t') dt'. \qquad (6)$$

The value $\sigma$ will be called the source deactivation coefficient. This coefficient indicates the rate at which the source loses its ability to generate aftershocks. We see that for $\sigma = \text{const}$ formula (5) coincides with Omori formula (1) up to notation.

Thus, we came to the conclusion that at the beginning of evolution, when $n(t) \gg n_\infty$, the logistic law (3) is practically indistinguishable from the Omori law (1) provided that $\sigma = \text{const}$. Our conclusion can be verified experimentally (see below).

### 4. New approach to the problem

"Measure what is measurable" / Galileo's advice

We begin to present a new approach to the study of aftershocks [12–16]. One of the Cartesian principles on which scientific thinking is based states that a problem must be clearly divided into simple and clearly visible parts. In the problem of aftershocks, we highlight the object, purpose and method of research. The new approach is radically different from that of Omori, Hirano and Utsu in all three parts.

We choose the earthquake source as the object of study, and not the aftershocks themselves. The goal is not a speculative selection of an empirical formula to describe the degradation of the frequency of aftershocks, but a search for the laws of evolution of the source after the formation of a main rupture in the continuity of rocks in it. The difference between the methods is the most dramatic. We took advantage of Galileo's powerful idea: observe, measure, and only on the basis of measurements draw theoretical conclusions instead of looking for speculative fitting formulas. Thus, the divergence between the two approaches is doctrinal.

Let us assume that from observations we know the dependence $n(t)$. First we will calculate the auxiliary function

$$g(t) = \frac{1}{n(t)} - \frac{1}{n_0}. \tag{7}$$

Now we will introduce a measure $\sigma(t)$, characterizing the current state of the source, as follows. Let's smooth the function $g(t)$, that is, average it over fast oscillations, and then differentiate it with respect to time:

$$\sigma(t) = \frac{d}{dt}\langle g(t) \rangle. \tag{8}$$

Here the angle brackets denote the smoothing procedure of the auxiliary function.

It was not by chance that we chose the symbol $\sigma$ to denote the measure of the state of the source. Taking into account (4), (7), (8), we can verify that Omori's law (1) is equivalent to the constancy of the source deactivation coefficient:

$$\sigma = \text{const}. \tag{9}$$

We can calculate $\sigma(t)$ based on measurements and check whether law (9) is satisfied.

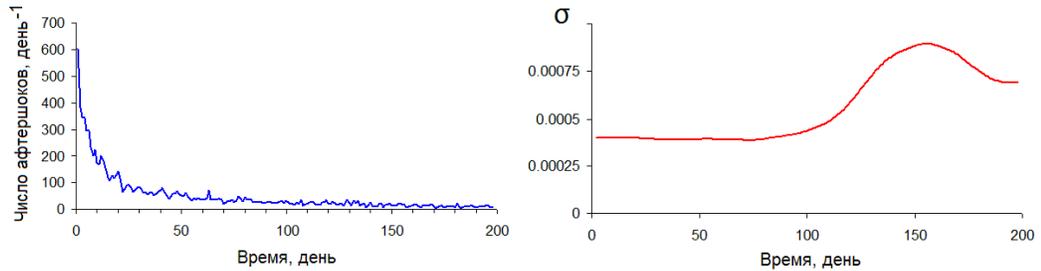

**Fig. 2.** Aftershock frequency (left) and source deactivation coefficient (right) after the strong earthquake in Southern California on January 17, 1994 ($M = 6.7$).

We have carried out critical testing of Omori's law within the framework of the new approach many times [12–16, 20–24]. One of the results is shown in Figure 2. The conclusion is that Omori's law is fulfilled, but only at the first stage of the evolution of the source. The period of time in which $\sigma = \text{const}$ was called the Omori epoch. The duration of the Omori epoch varies from case to case from several days to several months. There is a tendency for the duration to increase with increasing magnitude of the main shock. The source deactivation coefficient in the Omori epoch is lower, the higher the magnitude of the main shock of earthquake.

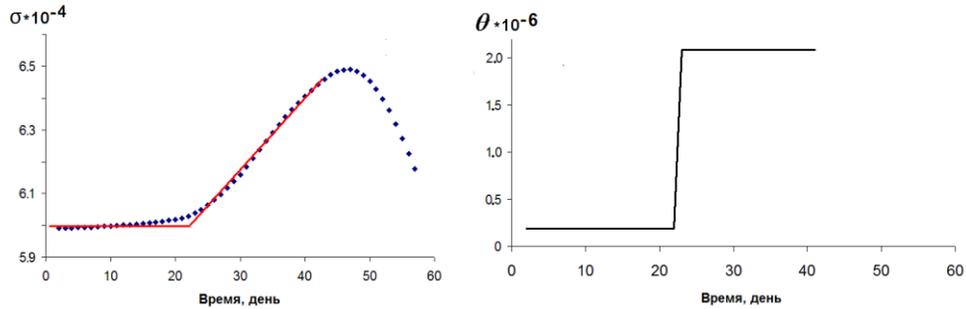

**Fig. 3.** Piecewise linear approximation of the deactivation coefficient (left) and its time derivative (right). The event occurred on November 23, 1984 in Northern California ($M = 6$).

At the end of the Omori epoch, the state of the source changes. Let's introduce the parameter $\theta = d\sigma/dt$. In the Omori epoch, $\theta = 0$. The transition to a new state is indicated by a sharp jump in the parameter $\theta$. This suggests that the end of the Omori epoch is accompanied by bifurcation of the earthquake source [25]. Figure 3 illustrates this [16].

We have not observed a single event in which Omori's hyperbolic law is satisfied throughout the entire evolution of aftershocks. Apparently, a hundred years ago, it was precisely this circumstance that served as a reason for Hirano to replace the Omori hyperbola with a power function (2).

So, Omori's law manifests itself only at the first stage of evolution. After that, bifurcation occurs. The transition from one source deactivation regime to a qualitatively different regime is abrupt in the sense that its duration is much less than the duration of the Omori epoch.

## 5. Discussion

We now understand the mistake of Hirano [2], who rejected the Omori law (1) and replaced it with the power law (2). Omori law in the form $\sigma = \text{const}$ manifests itself, but only for a limited period of time in the initial stage of the evolution of the source. At the end of this period, which we call the Omori epoch, the deactivation coefficient begins to change in a complex way over time. The methodological error made by Hirano 100 years ago [2] and later supported by Utsu [3–5] consists of a purely speculative selection of a fitting formula that best approximates the **entire** process of aftershock evolution. We discovered that evolution does not occur uniformly over time. Only at the first stage can the frequency of aftershocks be described by a simple formula, and this is a formula for a logistic curve, and not a formula for a hyperbola.

A methodological approach to the study of a natural phenomenon must be judged by its effectiveness. The new approach outlined above has proven to be very effective. We see that from the first steps of its development the method provided non-trivial information about the dynamics of the earthquake source. But the goal has not yet been achieved. We were unable to find the equation for the evolution of the source. One of the search directions is associated with the selection of the control parameter $\lambda$ and the order parameter $\varphi$ such that [16]

$$\frac{d\varphi}{dt} = \Gamma \varphi, \qquad (10)$$

with

$$\Gamma(\lambda, \varphi) = \Gamma_1(\lambda) + \Gamma_2(\varphi), \qquad (11)$$

where

$$\Gamma_1 = a(\lambda - \lambda_c), \quad \Gamma_2 = b\varphi - c\varphi^2. \qquad (12)$$

The values of $\lambda$, $\lambda_c$, $a$, $b$, $c$ are greater than zero. Let's set $\varphi = \theta$ and consider the case $\lambda < \lambda_c$. The equilibrium point $\theta = 0$ is stable and corresponds to the Omori epoch ($\sigma = \text{const}$). If the control parameter $\lambda$ increases smoothly and reaches a threshold of $\lambda_c$, then the order parameter increases abruptly from zero to a value of $\theta = b/c$. Qualitatively, this resembles the phenomenon of source bifurcation.

## 6. Conclusion

We performed a comparative analysis of two approaches to studying aftershocks. One of them was formed many years ago. Its essence comes down to the selection of empirical formulas that best approximate observational data. The new approach has begun to develop relatively recently. It differs from the traditional approach in the object, purpose and method of research. Briefly, the differences are as follows:

1. The focus is not on the aftershocks, but on the source of the earthquake.

2. The evolution of the source is studied experimentally, and not the degradation of the frequency of aftershocks.

3. Instead of a speculative selection of empirical formulas, the deactivation coefficient is measured, its variations are observed, and conclusions are drawn only on the basis of measurements and observations.

The new approach proved to be effective. It was possible to discover the existence of the Omori epoch and the phenomenon of bifurcation of a source. The problem of searching for a master equation that describes the evolution of an earthquake source remains unresolved.

*Acknowledgments*. I express my deep gratitude to B.I. Klain, A.D. Zavyalov and O.D. Zotov for many years of cooperation. Together with them, the foundations were laid for a new approach to the study of aftershocks.